# The ESO Science Archive

Martino Romaniello*[a]

[a]European Southern Observatory (ESO), Karl-Schwarschild-Strasse 2, 85748 Garching bei München, Germany

## ABSTRACT

The ESO Science Archive is the collection and access point of the data generated at ESO's La Silla Paranal Observatory, both raw and processed. It is a major contributor to ESO's science output, being used in about 4 out of 10 refereed articles with ESO data. In this paper, which is presented on behalf of the operations and development teams, we review its content, policies, user interfaces and impact.

## 1. INTRODUCTION

The ESO Science Archive is the collection and access point of the data generated at ESO's La Silla Paranal Observatory (LSP). The observatory comprises three sites in Northern Chile's Atacama region, namely Cerro La Silla (https://www.eso.org/public/teles-instr/lasilla), Cerro Paranal (https://www.eso.org/public/teles-instr/paranal-observatory) and the APEX telescope (https://www.eso.org/public/teles-instr/apex) on the Chajnantor plateau.

The ESO Science Archive stores all the raw science data from those facilities and the related calibrations. A growing selection of processed data is also available there, which provide direct access to frames on which science measurements can be readily performed. The archive home page is at: http://archive.eso.org/cms.html. User support and a knowledgebase database are provided at support.eso.org.

### 1.1 Data access policy

Access to the data holdings of the ESO Science Archive is regulated by policy (http://archive.eso.org/cms/eso-data-access-policy.html). In general terms, the Principal Investigators (PIs) of successful proposals for observing time on ESO telescopes, and their delegates, have exclusive access to their scientific data for the duration of a proprietary period, after which it is accessible to all users from the international community. The proprietary period is set by the ESO Director General and communicated at the time of proposing for observing time. It is typically of one year, but may depend on the observing programme type, as detailed in the policy (e.g. Public Survey raw data is not covered by proprietary protection). Processed data distributed via the SAF retain the same proprietary protections as the raw data they were derived from.

All data in the ESO archive retain ESO's copyright, and are distributed according to the terms of the Creative Commons Attribution 4.0 International license (CC BY 4.0, https://creativecommons.org/licenses/by/4.0). The use of ESO data, e.g. in publications, must be acknowledged (cf. http://archive.eso.org/cms/eso-data-access-policy.html).

## 2. DATA CONTENT AND ACCESS

### 2.1 Raw data

The footprint on the sky of the science raw data available in the ESO Science Archive as of June 2022 is shown in Figure 1.

The raw data can be queried by basic instrumental and scheduling criteria through a unified form (http://archive.eso.org/eso/eso_archive_main.html, see Figure 2). Specialized query forms for individual instrument, which expose many more detailed search parameters, are available at: http://archive.eso.org/cms/eso-data/instrument-specific-query-forms.html.

---

*martino.romaniello@eso.org; phone +49 89 32006565; www.eso.org

Once a user has selected the raw science data of interest, relevant calibration files can be associated automatically prior to downloading the data. The service is tuned to provide the calibrations as defined in the instrument's Calibration Plan. As an example, the one for the UVES high-resolution spectrograph can be found at: https://www.eso.org/sci/facilities/paranal/instruments/uves/doc/ESO_411894_Calibration_Plan_P110.pdf. At this stage, users can choose whether they want raw and/or pre-processed master calibrations, as well as to associate also nightlog information, such as weather conditions and notes form the observer (see Figure 3).

Raw data can, then, be processed to remove signatures from the telescope, instrument and the Earth atmosphere and to calibrate it in physical units using dedicated software tools (aka "pipelines", https://www.eso.org/sci/software/pipelines). At this point, the data is ready for extraction of the science signal.

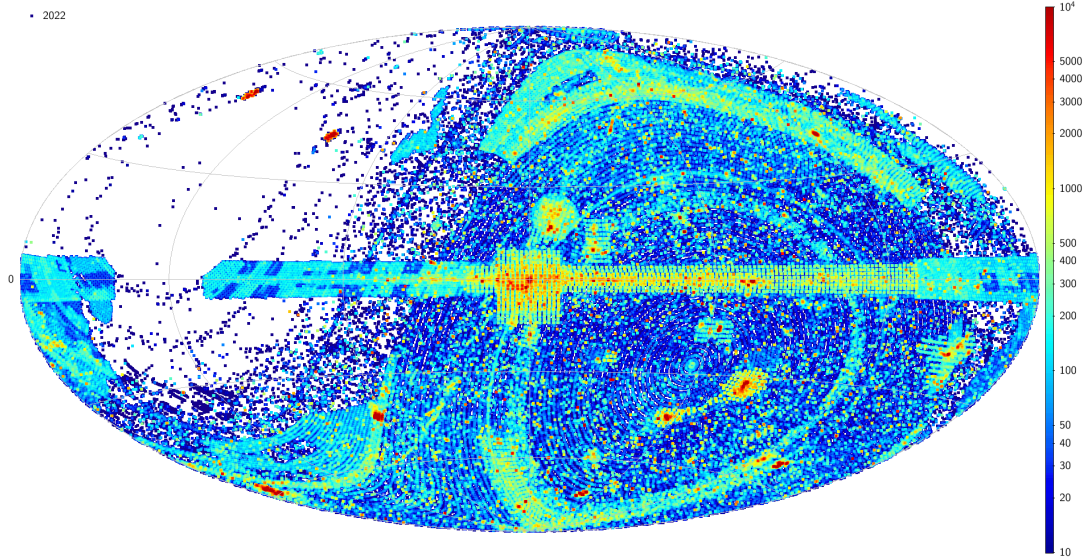

Figure 1 Sky coverage of the raw data content of the ESO Science Archive as of June 2022. Popular locations on the sky, such as the Galactic Centre or the Magellanic System, clearly stand out, as do the footprints of some of the Public Surveys. The few observations close to the Celestial North Pole are clearly the result of errors in metadata content.

Figure 2 The unified query form for raw data: http://archive.eso.org/eso/eso_archive_main.html.

Figure 3 Association of calibrations and nightlog information in action.

## 2.2 Processed data

To lower the barrier to using the data, the ESO Science Archive is filled with processed data. Via this route, users can download data that have already gone through most of the processing needed to extract the science signal, free from atmospheric and instrumental effects and calibrated in physical units. In most cases, data is directly ready for science analysis.

There are two basic sources of such processed data. One is external users doing the processing and returning the results to ESO. This is mandatory for observing programmes that require large, coordinated amount of telescope time, namely Public Surveys (https://www.eso.org/sci/observing/PublicSurveys/sciencePublicSurveys.html) and Large Programmes (https://www.eso.org/sci/observing/teles-alloc/lp.html). For those, generating data, both raw and processed, with a long-lasting legacy value is an important criterion in the programme selection process. Voluntary contributions from individual users are encouraged, although not mandatory. The processing is tailored to the science case(s) of the originating observing programmes, and results are often described and used by the team in refereed publications. Typical data products include calibrated deep and/or mosaiced images and data cubes, stacked spectra, and flux maps. In several cases, these are used to generate source catalogues, which are, then, also published through a dedicated channel in the archive. Source catalogues are the highest level of processed data, in that they contain the physical parameters of the celestial sources.

The other main source of processed data into the Science Archive is by ESO doing the processing directly, for archival ingestion and dissemination. Here, the data histories of instruments, or instrument modes, are processed consistently and as completely as possible. By its very nature, processing is not tailored to any specific science case, but is focused on removing the instrumental and atmospheric signatures and on calibrating in physical units.

The impact of archival processed data is discussed in Section 5 below.

## 3. DATA PRODUCT STEWARDSHIP: THE PHASE 3 PROCESS

At the time of writing, the ESO Science Archive contains more than 3.5 million science files from nearly 70 data collections, covering virtually all data types and observational techniques enabled by the slew of instruments that ESO operates. Before being ingested into the archive, they undergo an auditing process for completeness, compliance, consistency, and documentation. This process is called Phase 3, to reflect the fact that it closes the loop after the solicitation and handling of observing proposals (Phase 1) and the preparation of the observations themselves (Phase 2).

To ensure data consistency and accessibility throughout this broad variety of archive holdings, the Phase 3 process enforces the use of a data standard. The ESO Science Data Product Standard (http://www.eso.org/sci/observing/phase3/p3sdpstd.pdf) is an interface document that defines the data and metadata format and information for the various data types (images, cubes, interferometric visibilities, catalogues, …). It defines how to encode the level of calibration, scientific quality, originating files (provenance), ancillary data and product versioning. It is continuously evolving to reflect the evolving data landscape, e.g. to incorporate new data types os user cases.

The ESO Science Data Product Standard is based on accepted Virtual Observatory (VO) standards, so that the ESO data are made interoperable with other VO resources. It is a fundamental tool to make each data collection findable within the other ESO data, and in the general global data scenario.

## 4. THE ESO ARCHIVE SCIENCE PORTAL

An important driver for the metadata curation and homogenization ensuing from the Phase 3 process is that data can be presented and queried uniformly across data collections, independently of their origins and specificities. We have built two ways to browse the processed data in the ESO Science Archive, namely a web interface and programmatic access.

The archive web interface offers a low barrier to its use by presenting the data and metadata in an intuitive graphical interface (http://archive.eso.org/scienceportal/home). Query parameters are represented by facets that have the dual function of visually expressing and representing the user choices. Query results are, then, rendered on the backdrop of the celestial sphere and included in a tabular form that summarizes their main characteristics (see Figure 4). Upon request, individual datasets can be explored in great detail individually, including through previews that are customized by data type. An example for MUSE stacked cubes is shown in Figure 5. Given the underlying compatibility with VO protocols, the results can be seamlessly sent to VO-aware tools, like, e.g., TOPCAT or Aladin.

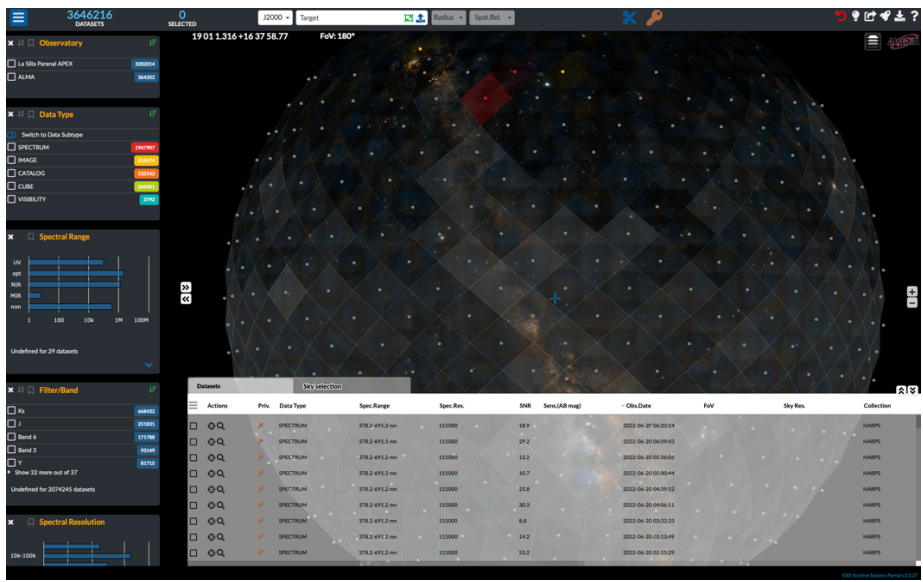

Figure 4 The landing page of the web interface to the ESO Archive Science Portal at http://archive.eso.org/scienceportal/home. See the text for a description of the different elements.

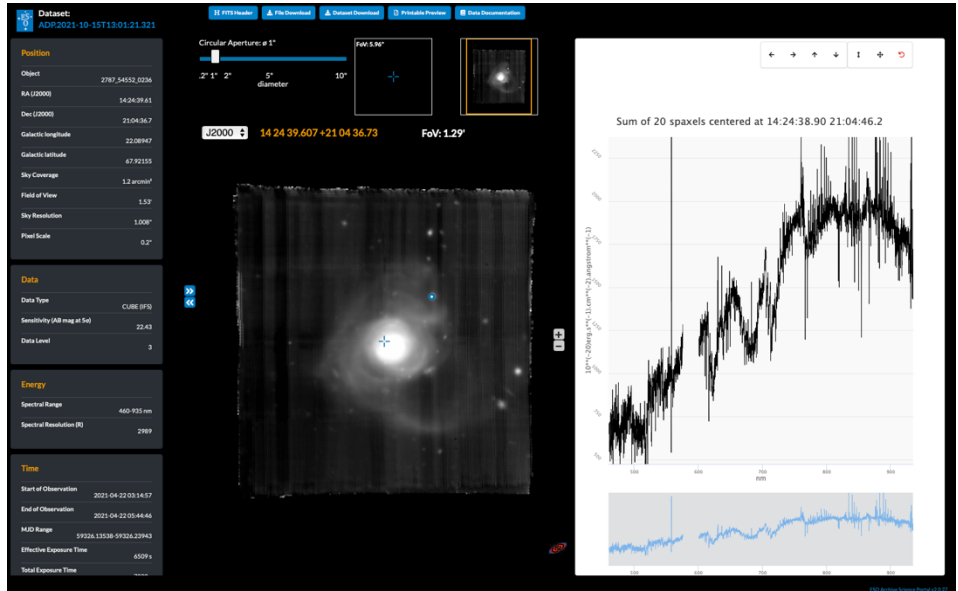

Figure 5 Example of the web page where individual datasets can be explored in detail. In this case, the file is a stack of MUSE datacubes. The dataset can be rendered in 2D with a custom level of zoom. By clicking on a specific position, the 1D extracted and calibrated spectrum at that location is presented in a dedicated spectral viewer (right-hand side panel in the figure).

Repetitive or otherwise particularly demanding tasks can be coded and automatized by means of the provided programmatic access tools (http://archive.eso.org/programmatic). They make extensive use of VO protocols, thus ensuring interoperability and a high level of standardization. Extensive documentation and examples are provided to guide users in their first steps, and throughout more complex cases.

Finally, a dedicated interface is available at https://www.eso.org/qi to query source catalogue data. Once users select the catalogue they are interested in, they have the possibility to constrain the search on any of the catalogue columns specific to it.

Online documentation, examples, including in the form of software scripts, and a user community forum are available at esocommunity.userecho.com.

## 5. THE IMPACT OF THE ESO SCIENCE ARCHIVE

As shown in Figure 6, where the cumulative number of unique archive users is shown as a function of time for the different types of data holdings, all of them are in great demand. It is interesting to note that the increase in the adoption of processed data has not come at the expenses of the access to raw data, just like the fast increase of users of data processed by ESO has not hindered the need for data generated externally. The different types of data are, then, highly complementary.

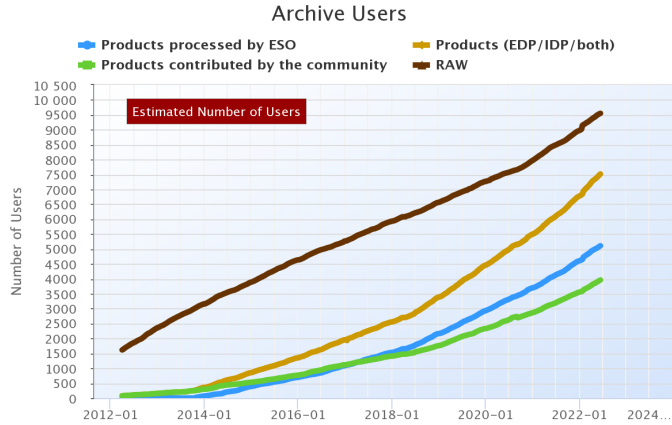

Figure 6 The cumulative number of unique users as a function of time for the different data type holding described in the text: raw (brown line), products generated at ESO (blue line) and products contributed by the community (green line). The dark yellow line represents the number of unique users of either type of processed data.

To conclude, in Figure 7 we report the contribution of the Science Archive to the science output of ESO, expressed in terms of the fraction of refereed papers using ESO data that made use of the archive. As it can be seen, it has a clear upwards trend since the inception of the ESO Science Archive in its current form at the end of the 1990s. Currently, about 4 papers out of 10 make use of the ESO Science Archive, a very satisfying result indeed!

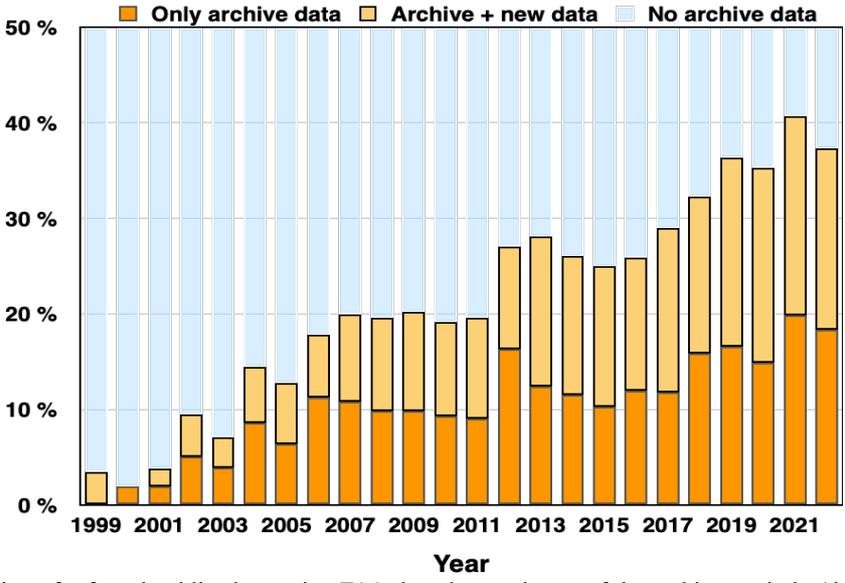

Figure 7 The fraction of refereed publications using ESO data that made use of the archive, entirely (dark orange bars) or partially (light yellow bars). Source: ESO Telescope Bibliography (https://telbib.eso.org).